\documentclass{aa}
\usepackage{graphicx}
\usepackage{natbib}
\usepackage{txfonts}
\begin{document}

\title{On the evolution of giant radio halos and their connection with
cluster mergers}

  \subtitle{}
   
\author{G. Brunetti\inst{1}  \and R. Cassano\inst{1}  
\and K. Dolag\inst{2}  \and G. Setti\inst{3}
}
\institute{INAF- Istituto di Radioastronomia, via P. Gobetti 101,
I--40129 Bologna, Italy
\and
Max-Planck-Institut f\"ur strophysik, Karl-Schwarzschild
Strasse 1, D-85741 Garching bei M\"unchen, Germany
\and
Dip. Astronomia, Universita' di Bologna, via Ranzani 1,
I--40127 Bologna, Italy
}

\date{}

\authorrunning{G. Brunetti et al.}
\titlerunning{Radio--X-ray correlation of giant Radio Halos}

\abstract{}{
Giant radio halos are diffuse, Mpc-scale, synchrotron sources located in
the central regions of galaxy clusters and provide the most 
relevant example of cluster non-thermal activity.
Radio and X-ray surveys allow to investigate the statistics of 
radio halos and may contribute to constrain the origin of these sources and 
their evolution.}
{We investigate the distribution of clusters in the plane 
X--ray (thermal, $L_X$) vs synchrotron (non-thermal, $P_{1.4}$) 
luminosity, where clusters hosting giant radio halos trace the 
$P_{1.4}$--$L_X$ correlation and clusters without radio halos
populate a region that is well separated from that spanned by
the above correlation.
The connection between radio halos and cluster mergers suggests that
the cluster Mpc-scale synchrotron emission is amplified during
these mergers and then suppressed when clusters become more
dynamically relaxed.}
{In this context, by analysing the distribution in the $P_{1.4}$--$L_X$ plane
of galaxy clusters from X-ray selected samples with adequate radio follow up, 
we constrain the typical time-scale of evolution of diffuse radio emission
in clusters and discuss the implications for the origin of radio halos.}
{We conclude that cluster synchrotron 
emission is suppressed (and amplified) in a time-scale significantly
smaller than 1 Gyr. We show that this constraint appears difficult to
reconcile with the hypothesis that the halo's radio power is suppressed
due to dissipation of magnetic field in galaxy clusters.
On the other hand, in agreement with models where turbulent acceleration 
plays a role, present constraints
suggest that relativistic electrons are accelerated in Mpc-scale regions, 
in connection with cluster mergers and for a time-interval of about
1 Gyr, and then they cool in a relatively small time-scale, when 
the hosting cluster becomes more dynamically relaxed.}


 \maketitle

\section{Introduction}\label{intro}

Radio observations of galaxy clusters unveil the presence of
relativistic particles and magnetic fields in the intracluster 
medium (ICM) through the detection of 
diffuse Mpc-scale synchrotron emission in the form of {\it radio halos}
and {\it radio relics} (e.g., Ferrari et al. 2008 for recent review).

\noindent
Giant radio halos provide the most spectacular evidence of non-thermal
phenomena in the ICM. They are giant diffuse radio sources located at
the centre of galaxy clusters and extending similarly to the hot ICM;
remarkably they are always found in clusters with evidence for ongoing
mergers (e.g. Buote 2001; Govoni et al. 2004; Venturi et al. 2008).
These halos prove that mechanisms of in situ particle acceleration or
injection are active in the ICM since the diffusion time necessary to
the radio emitting electrons to cover Mpc scales is much longer than
their radiative lifetime (e.g., Jaffe 1977).

\noindent
Correlations between the radio power at 1.4 GHz of giant radio halos
($P_{1.4}$) and their physical size, and between $P_{1.4}$ and
the X-ray luminosity ($L_X$) and temperature of the hosting clusters
have been found and discussed in the literature (e.g. Liang et al. 2000;
Bacchi et al. 2003; Cassano et al. 2006,07; Brunetti et al. 2007;
Rudnick et al. 2009). These correlation suggest that gravity provides
the reservoir of energy to generate the non-thermal components responsible
for the emission from the ICM.

Mergers drive shocks and turbulence in the ICM that may lead to
the amplification of the magnetic fields (e.g., Dolag et al. 2002;
Subramanian et al. 2006; Ryu et al. 2008) and to the acceleration
of high energy particles (e.g., En\ss lin et al. 1998; Roettiger et al.~1999;
Sarazin 1999; Blasi 2001; Brunetti et al. 2001, 2004;
Petrosian 2001; Fujita et al. 2003; Ryu et al. 2003;
Hoeft \& Br\"uggen 2007; Brunetti \& Lazarian 2007).
More specifically, 
extended and fairly regular diffuse radio emission may be produced
by secondary electrons injected during proton-proton collisions, since 
relativistic protons can diffuse on large scales (hadronic or seconday 
models; e.g., Dennison 1980; Blasi \& Colafrancesco 1999), or by 
relativistic electrons re-accelerated in situ by MHD turbulence generated 
in the ICM during cluster-cluster mergers (re-acceleration models; e.g., 
Brunetti et al. 2001; Petrosian 2001).
Observations provide support to the idea that turbulence may play a role in 
the particle acceleration process (e.g., Brunetti 2008; Ferrari et al. 2008; 
Cassano 2009 for recent reviews).
Low frequency radio observations (e.g., with LOFAR, 
LWA) and high energy observations (with FERMI) are expected to set crucial 
constraints.

\begin{figure*}
\begin{center}
\includegraphics[width=7cm]{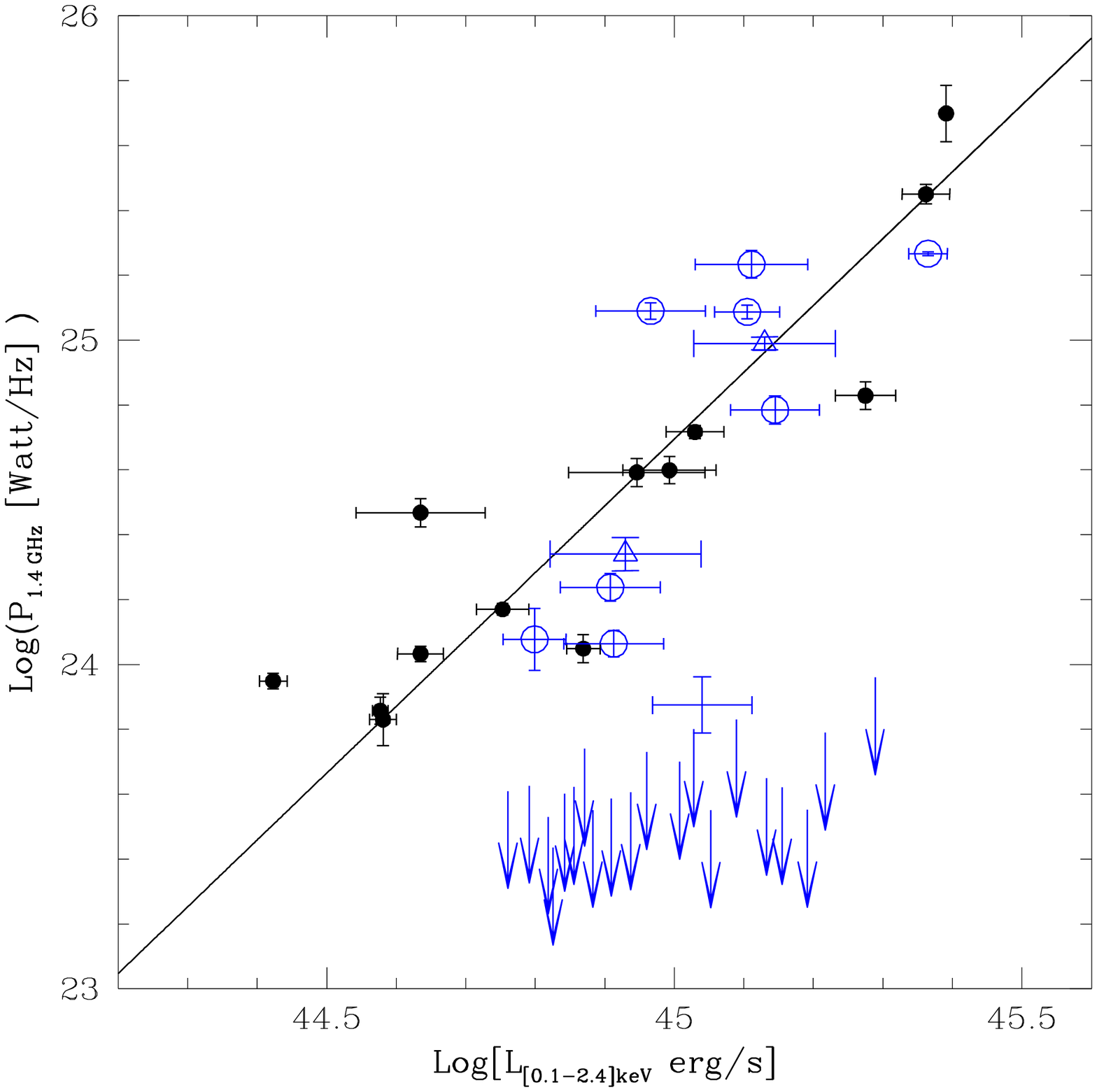}
\includegraphics[width=7cm]{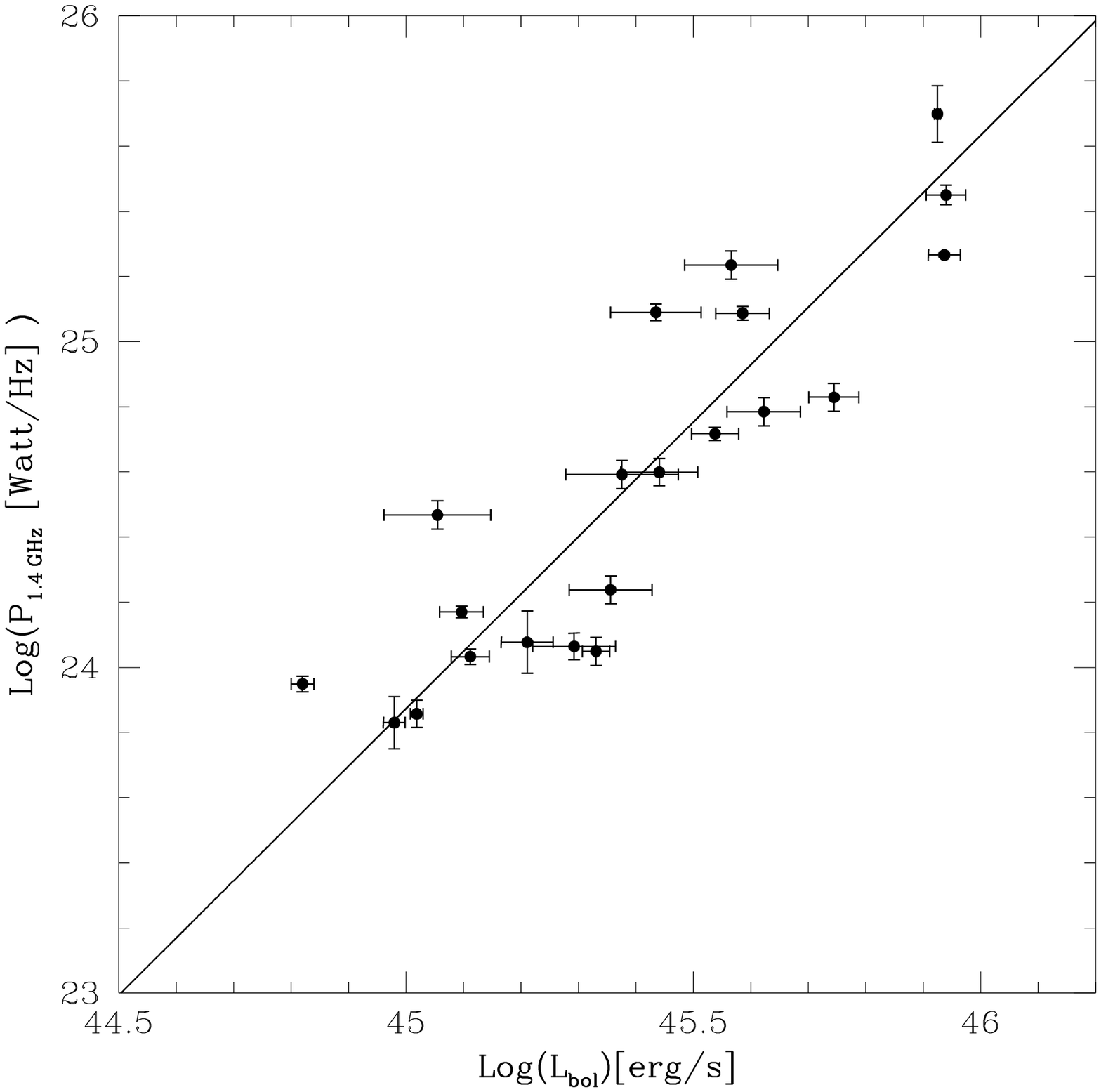}
\caption{{\bf Left} panel: distribution of GMRT galaxy clusters (blue)
and of other radio-halo clusters from the literature (filled black symbols) 
in the $P_{1.4}$--0.1-2.4 keV luminosity plane (Table 1). 
Empty circles mark giant radio halos from the GMRT sample, empty triangles
mark the two mini-halos in cool-core clusters from the GMRT sample, the 
cross marks the position of RXJ1314, and
arrows mark upper limits for GMRT clusters with no evidence of Mpc-scale
radio emission. The solid line gives the best fit to the distribution of
giant radio halos (BCES Bisector, Table 2).
{\bf Right} panel: distribution of giant radio halos (GMRT + literature,
Table 1) in the $P_{1.4}$--bolometric X-ray luminosity plane. 
The solid line gives the best fit to the distribution of giant radio halos 
(BCES Bisector, Table 2).
}
\end{center}
\end{figure*}

In this paper we study the distribution of X--ray selected
galaxy clusters in the $P_{1.4}$--$L_X$ plane providing
an extension of a previous work (Brunetti et al.~2007).
More specifically, 
we discuss constraints on the relevant time-scales of the evolution of
magnetic fields and emitting particles in the ICM and their
consequence on the origin of radio halos.

In Sect.~2 we discuss the distribution of 
X--ray luminous galaxy clusters in the $P_{1.4}$--$L_X$ plane
and the $P_{1.4}$--$L_X$ correlation traced by radio halos. In Sect.~3
we discuss the connection between mergers and radio halos and their
evolution driven by these mergers.
In Sects.~4 and 5 we constrain the evolution time-scale of radio halos
and compare our results with model expectations, respectively.
In Sect.~6 we give our Conclusions.

\noindent
$H_o$=70 km s$^{-1}$ Mpc$^{-1}$, $\Omega_m$=0.3, $\Omega_{\Lambda}$=0.7 are
adopted throughout the paper.

\section{The Radio -- Lx correlation \& cluster bi-modality}

The ``GMRT Radio Halo Survey'' (Venturi et al. 2007, 2008)
has provided a significant step to a statistically solid
exploration of the properties of radio halos through a large observational
project carried out with the Giant Metrewave Radio Telescope
(GMRT, Pune-India) at 610 MHz. 
This pointed-radio survey completed the radio follow up of a complete sample
of 50 X-ray luminous ($L_X\geq5\cdot10^{44}$ erg/s) galaxy clusters in the
redshift range $0.2-0.4$ (taken from the REFLEX, Boehringer et al. 2004 and
the extended BCS, Ebeling et al. 1998, 2000 catalogues) through high 
sensitivity observations of 34 clusters with no 
radio information. 
Large scale synchrotron emission at level of presently known radio halos 
was found only in $\sim30\%$ of the selected 
(X-ray luminous and massive) clusters (Brunetti et al. 2007; 
Venturi et al. 2008), with evidence that
this fraction depends on cluster X--ray luminosity (Cassano et al. 2008).

Figure 1 shows the distribution of GMRT galaxy clusters (blue) in the 
$P_{1.4}$ -- $L_X$ plane (Left : 0.1-2.4 keV luminosity; Right : bolometric 
luminosity), together with that of clusters hosting giant radio halos (from 
the literature, Table 1): giant 
radio halos trace the $P_{1.4}$ -- $L_X$ 
correlation, clusters with no large scale radio emission 
populate the region of the radio 
upper limits that is well separated from that spanned by radio halos.

\begin{table*}
\caption{Radio and X-ray properties of clusters used in this paper:
giant radio halos (first sector), mini-halos in cool core clusters
(second sector), small scale halos (tirdth sector), clusters with no
Mpc-scale radio emission (last sector). GMRT clusters are reported in
bold.
In Col.(1): Cluster name. Col.(2): Cluster redshift. 
Col.(3): X-ray luminosity in the energy range $[0.1-2.4]$ keV in unit 
of $h_{70}^{-2}\,10^{44}$ erg/s.
Col.(4): Bolometric X-ray luminosity in the energy range $[0.01-40]$ keV 
in unit of $h_{70}^{-2}\,10^{44}$ erg/s.
Col.(5): Radio power at $1.4$ GHz in unit of $h_{70}^{-2}\, 10^{24}$ Watt/Hz. 
Col.(6) References: 1 = Boehringer et al 2004, 2 = Ebeling et al 1998,
3 = Ebeling et al 1996, 4 = Tsuru et al 1996, 5 = Ebeling et al 2007,
6 = Ebeling et al 2000, 7 = Liang et al. 2000, 8 = Feretti et al. 2001,
9 = Govoni et al. 2001, 10 = Bacchi et al 2003, 11 = Giovannini \& Feretti
2000, 12 = Dallacasa et al 2009, 13 = Clarke \& Ensslin 2006,
14 = Brentjens 2008, 15 = Feretti 2002, 16 = Kim et al. 1990,
17 = Deiss et al. 1997, 18 = Govoni et al. 2005, 19 = Venturi et al 2007,
20 = Giacintucci et al 2009, 21 = van Weeren et al 2009, 
22 = Bonafede et al 2009, 23 = Cassano et al 2008b, 24 = Giacintucci 2007
(integrating the diffuse radio emission between the two radio relics).
(*) $P_{1.4}$ of A209 and RXJ1314 is estimated from that at 610 MHz by adopting
$\alpha =1.2$.}
\begin{center}
\begin{tabular}{llllllllllll}
\hline
\hline
cluster's  &  z &     $L_X$          & $L_{bol}$         &   $P_{1.4}$ & 
Ref      \\
  name     &    &  [$10^{44}$ erg/s ]& [$10^{44}$ erg/s ]& [$10^{24}$
  Watt/Hz] & \\
\hline
\hline
1E50657-558  & 0.2994  &  $23.03\pm 1.81$  &   $87.10\pm 6.82$ &
$28.21\pm 1.97$ & 1, 7  \\
{\bf A2163}  & 0.2030  &  $23.17\pm 1.48$  &   $86.50\pm 5.52$ &
$18.44\pm 0.24$ & 1, 8 \\
{\bf A2744}  & 0.3080  &  $12.92\pm 2.41$  &   $36.81\pm 6.88$ &
$17.16\pm 1.71$ & 1, 9 \\
{\bf A2219}  & 0.2280  &  $12.73\pm 1.37$  &   $38.55\pm 4.05$ &
$12.23\pm 0.59$ & 2, 10 \\
CL0016+16    & 0.5545  &  $18.83\pm 1.88$  &   $55.59\pm 5.56$ &
~$\:6.74\pm 0.67$ & 4, 11\\
A1914        & 0.1712  &  $10.71\pm 1.02$  &   $34.51\pm 3.28$ & ~$\:5.21\pm
0.24$ & 3, 10 \\
A665         & 0.1816  & ~$\:9.84\pm 1.54$ &   $27.61\pm 4.32$ & ~$\:3.98\pm
0.39$ & 2, 11 \\
A520         & 0.2010  & ~$\:8.83\pm 1.99$ &   $23.77\pm 5.34$ & ~$\:3.91\pm
0.39$ & 2, 10 \\
{\bf A521}   & 0.2475  & ~$\:8.18\pm 1.36$ &   $19.63\pm 3.26$ & ~$\:1.16\pm
0.11$ & 1, 12 \\   
A2254        & 0.1780  & ~$\:4.32\pm 0.92$ &   $11.35\pm 2.43$ & ~$\:2.94\pm
0.29$ & 3, 10 \\
A2256        & 0.0581  & ~$\:3.81\pm 0.17$ &  ~$\:9.54\pm 0.43$& ~$\:0.68\pm
0.12$ & 3, 13, 14 \\
{\bf A773}   & 0.2170  & ~$\:8.10\pm 1.35$ &   $22.70\pm 3.78$ & ~$\:1.73\pm
0.17$ & 2, 9 \\
A545         & 0.1530  & ~$\:5.66\pm 0.49$ &   $12.50\pm 1.09$ & ~$\:1.48\pm
0.06$ & 1, 10 \\
A2319        & 0.0559  & ~$\:7.40\pm 0.40$ &   $21.42\pm 1.16$ & ~$\:1.12\pm
0.11$ & 3, 15 \\
{\bf A1300}  & 0.3071  &  $13.96\pm 2.05$ &   $41.97\pm 6.17$ & ~$\:6.09\pm
0.61$ & 1, 15 \\
A1656 (Coma) & 0.0231  & ~$\:3.77\pm 0.10$ &   $10.44\pm 0.28$ &
~$\:0.72^{+0.07}_{-0.04}$ & 3, 16, 17 \\
A2255        & 0.0808  & ~$\:2.65\pm 0.12$ &  ~$\:6.61\pm 0.30$& ~$\:0.89\pm
0.05$ & 3, 18 \\
A754         & 0.0535  & ~$\:4.31\pm 0.33$ &   $12.94\pm 0.99$ & ~$\:1.08\pm
0.06$ & 3, 10 \\
{\bf A209}   & 0.2060  & ~$\:6.29\pm 0.65$ &   $16.26\pm1.69$  & ~$\:1.19\pm
0.26$ & 1, 19* \\
{\bf RXJ2003}& 0.3171  & ~$\:9.25\pm 1.53$ &   $27.23\pm4.95$  &  $12.30\pm
0.71$ & 1, 20 \\
MACS J0717   & 0.5548  &  $24.6\pm 0.3$    &   $84.18\pm1.01$  &  $50.0\pm
10.0$ & 5, 21, 22 \\
\hline
{\bf A2390} & 0.228 & $13.49\pm 3.16$ & & ~$\:9.77 \pm 0.45$ & 2, 10 \\
{\bf Z7160} & 0.258 & ~$\:8.51\pm 2.12$ & & ~$\:2.19 \pm 0.26$ & 2, 23 \\
\hline
{\bf RXJ1314}& 0.2439 & $10.96 \pm 1.81$ & & ~$\:0.75 \pm 0.15$ & 1, 19*, 24* \\
\hline
{\bf A2697}        & 0.2320  &   ~$\:6.88 \pm 0.85$  &   &  $<$0.40 & 1, 25 \\
{\bf A141}         & 0.2300  &   ~$\:5.76 \pm 0.90$  &   &  $<$0.36 & 1, 25 \\
{\bf A3088}        & 0.2537  &   ~$\:6.95 \pm 1.20$  &   &  $<$0.42 & 1, 25 \\
{\bf RXJ1115.8}    & 0.3499  &   $13.58\pm 2.99$     &   &  $<$0.45 & 1, 25 \\
{\bf S780}         & 0.2357  &   $15.53\pm 2.80$     &   &  $<$0.36 & 1, 25 \\
{\bf RXJ1512.2}    & 0.3152  &   $10.19\pm 1.76$     &   &  $<$0.63 & 1, 25 \\
{\bf A2537}        & 0.2966  &   $10.17\pm 1.45$     &   &  $<$0.50 & 1, 25 \\
{\bf A2631}        & 0.2779  &   ~$\:7.57 \pm 1.50$  &   &  $<$0.39 & 1, 25 \\
{\bf A2667}        & 0.2264  &   $13.65\pm 1.38$     &   &  $<$0.42 & 1, 25 \\
{\bf RXJ0027.6}    & 0.3649  &   $12.29\pm 3.88$     &   &  $<$0.68 & 6, 25 \\
{\bf A611}         & 0.2880  &   ~$\:8.86 \pm 2.53$  &   &  $<$0.40 & 6, 25 \\
{\bf A781}         & 0.2984  &   $11.29\pm 2.82$     &   &  $<$0.36 & 2, 25 \\
{\bf Z2089}        & 0.2347  &   ~$\:6.79 \pm 1.76$  &   &  $<$0.27 & 2, 25 \\
{\bf Z2701}        & 0.2140  &   ~$\:6.59 \pm 1.15$  &   &  $<$0.42 & 2, 25 \\
{\bf A1423}        & 0.2130  &   ~$\:6.19 \pm 1.34$  &   &  $<$0.41 & 2, 25 \\
{\bf Z5699}        & 0.3063  &   ~$\:8.96 \pm 2.24$  &   &  $<$0.54 & 6, 25 \\
{\bf Z5768}        & 0.2660  &   ~$\:7.47 \pm 1.66$  &   &  $<$0.36 & 6, 25 \\
{\bf Z7215}        & 0.2897  &   ~$\:7.34 \pm 1.91$  &   &  $<$0.55 & 6, 25 \\
{\bf RXJ1532.9}    & 0.3450  &   $16.49\pm 4.50$     &   &  $<$0.62 & 2, 25 \\
{\bf RXJ2228.6}    & 0.4177  &   $19.44\pm 5.55$     &   &  $<$0.91 & 6, 25 \\
\hline     
\hline
\label{RH}
\end{tabular}
\end{center}
\end{table*}

\begin{table*}
\begin{center}
\begin{tabular}{c|c|c|c|c|c|c|c}
\hline
\hline
   & & A & $\sigma_A$ & b & $\sigma_b$ & $\sigma_{Log(P_{1.4})}$ &
   $\sigma_{Log(L_X)}$ \\
\hline
$P_{1.4}$--$L_{0.1-2.4}$ & BCES Bisector & 0.195 & 0.060 & {\bf 2.06} &
   {\bf 0.20} & 0.28 & 0.14 \\
   & bootstrap     & 0.192 & 0.063 & 2.08 & 0.22 &  &      \\
\hline
$P_{1.4}$--$L_{0.1-2.4}$ & BCES Orthogonal & 0.204 & 0.062 & 2.21 & 0.23
   &      &      \\
   & bootstrap        & 0.202 & 0.067 & 2.23 & 0.27
   &      &      \\
\hline
$P_{1.4}$--$L_{bol}$    & BCES Bisector & 0.077 & 0.057 & {\bf 1.76} & 
{\bf0.16} & 0.26 & 0.15 \\
   & bootstrap     & 0.075 & 0.059 & 1.78 & 0.18 & &      \\
\hline
$P_{1.4}$--$L_{bol}$ & BCES Orthogonal & 0.077 & 0.059 & 1.86 &
	        0.19 &      &      \\
   & bootstrap       & 0.075 & 0.062 & 1.88 & 0.22 &      &      \\
\hline
\hline
\end{tabular}
\end{center}
\caption{Fitting parameters for the correlations between
$P_{1.4}$--$L_{0.1-2.4}$ and $P_{1.4}$--$L_{bol}$ of giant
radio halos (see text).}
\end{table*}

\noindent
The distribution of giant halos across the correlation
is significantly broader
than the typical error bars in their measured radio and X-ray luminosities
thus implying a possible intrinsic scatter in the correlation.
For this reason, as in Cassano et al.(2006), in our analysis we use 
the linear regression algorithm by Akritas \& Bershady (1996) that 
indeed accounts for both intrinsic scatter and measured errors 
in both variables.
The fits have been performed in the form :

\begin{equation}
{\rm Log} ( P_{1.4} ) - Y
= A + b \left[ {\rm Log} ( L_{X} ) - X \right]
\end{equation}

\noindent
where $P_{1.4}$ is in W/Hz, $L_X$ is in erg/s,
$Y$=24.5, and $X$=45 and 45.4 in the case of 
the 0.1-2.4 keV luminosity (Fig.~1 Left) and bolometric X-ray luminosity
(Fig.~1 Right), respectively.
The best-fit normalizations and slopes of the correlations for giant 
radio halos,
and the measured scatters across the correlation are given 
in Table 2.
We report best fits obtained from BCES-bisector and
orthogonal approaches and from their bootstraps (10000 bootstrap
resamplings). 
The BCES bisector approach treats the variables symmetrically and is
recommended for scientific problems where the goal is to estimate
relationships between the variables and for a comparison with 
theory (e.g., Isobe et al.~1990).

Regardless of the nature of the scatter of the datapoint across the
correlation, we point out that
present data allow us to fairly constrain the slope of the correlations
\footnote{These slopes are consistent with those 
found in previous papers (e.g., Bacchi et al.~2003; Cassano et al.~2006).
Kushnir et al.(2009) recently found significantly flatter slopes,
however they do not fit the data with their 
errors {\it measured} in both variables but {\it assuming} 
an error in $P_{1.4}$ equal to the scatter of data points 
across the correlation.}.

\section{Evolution of radio halos and connection with cluster mergers}

Correlations between radio and thermal properties
in galaxy clusters may be explained by both hadronic and
turbulent-acceleration models with the observed 
slopes that can be reproduced provided that also the magnetic field
scales with cluster mass, temperature or luminosity (e.g. Dolag \&
Ensslin 2000; Miniati
et al.~2001; Cassano et al.~2006; Dolag 2006).

The new result from the high sensitivity
data of the ``GMRT Radio Halo Survey''
is the {\it bi-modal} behaviour in Fig.~1, with radio-halo 
clusters and clusters without radio halos clearly separated 
(Brunetti et al.~2007).
It is not obvious to understand how two well separated classes of 
galaxy clusters can be generated in the context of the 
hierarchical process of large scale structure formation. 
Indeed clusters with similar thermal properties (mass, X-ray luminosity,
{\ldots}) are expected to have a similar probability to host radio halos.
In this case 
the observed differences in terms of cluster non-thermal properties 
should be understood by assuming different evolutionary stages of 
these clusters.

\noindent
Clusters in the GMRT sample are selected with 
similar X-ray luminosity ($\approx$mass) and redshift. 
The radio halo -- merger connection (e.g. Venturi et
al.~2008) may suggest that the difference between giant radio halo 
and ``radio quiet'' clusters is due to their dynamics.
Thus regardless of the details of the mechanisms that generate
radio halos, we shall assume the following evolutionary cycle :

\begin{itemize}

\item{\it i)} galaxy clusters host giant radio halos for a period of time,
in connection with cluster mergers, and populate the 
$P_{1.4}$--$L_X$ correlation;

\item{\it ii)} at later times,
when clusters become dynamically relaxed, the Mpc-scale
synchrotron emission 
is gradually suppressed and clusters populate 
the region of the upper limits.

\end{itemize}

In this case, when restricting to clusters of the GMRT complete sample, 
Figure 1 provides a fair statistical sampling of the 
evolutionary
flow of X--ray luminous clusters in the $P_{1.4}$--$L_X$ plane
at z=0.2-0.4.

\noindent
Radio-halo clusters, always dynamically
disturbed systems, must be the ``youngest'' systems, where an ongoing
merger, leading to their formation (or accretion of a sizable fraction
of their mass), is still supplying energy to maintain the synchrotron
emission.

\begin{figure*}
\begin{center}
\includegraphics[width=7cm]{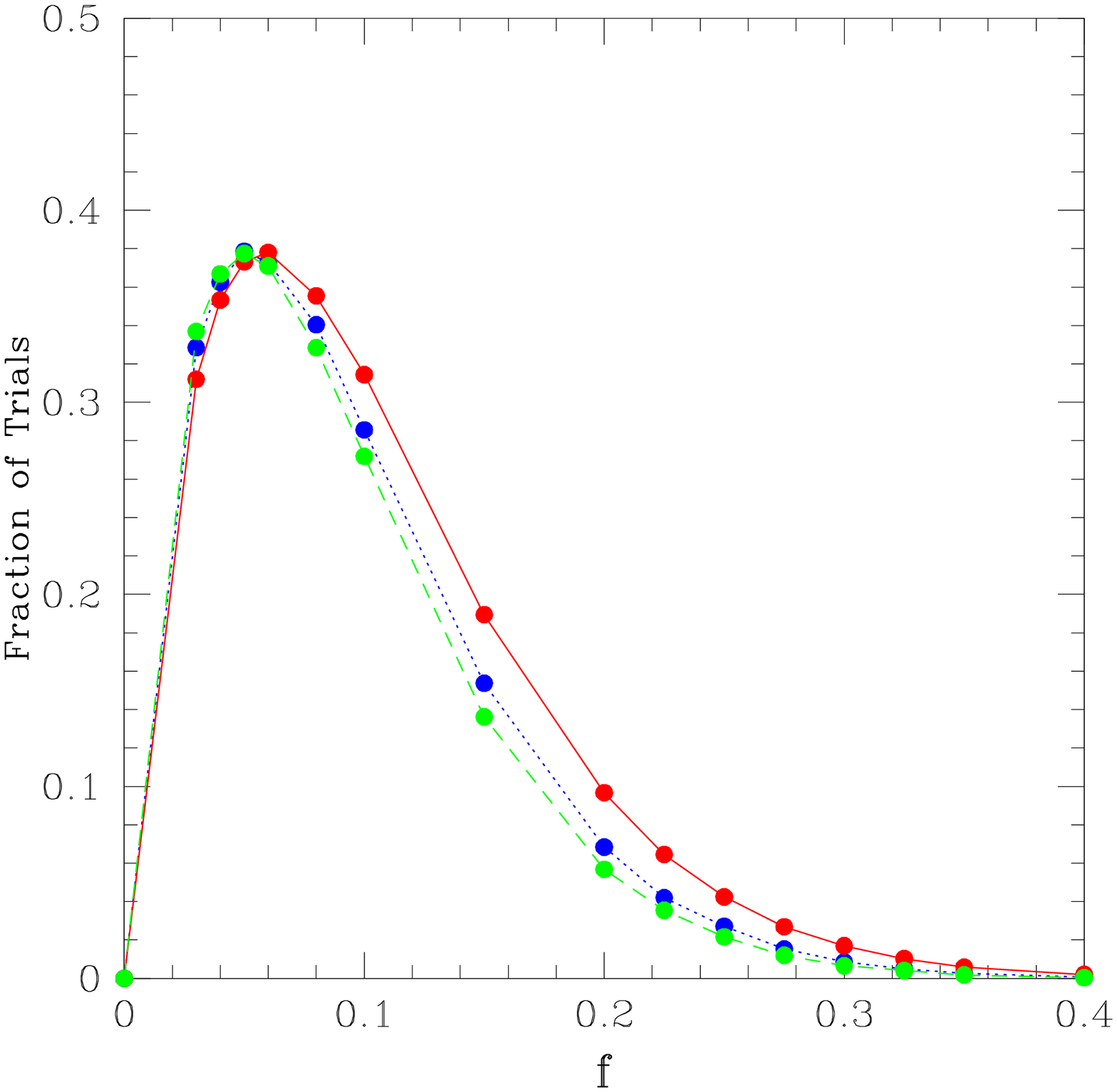}
\includegraphics[width=7cm]{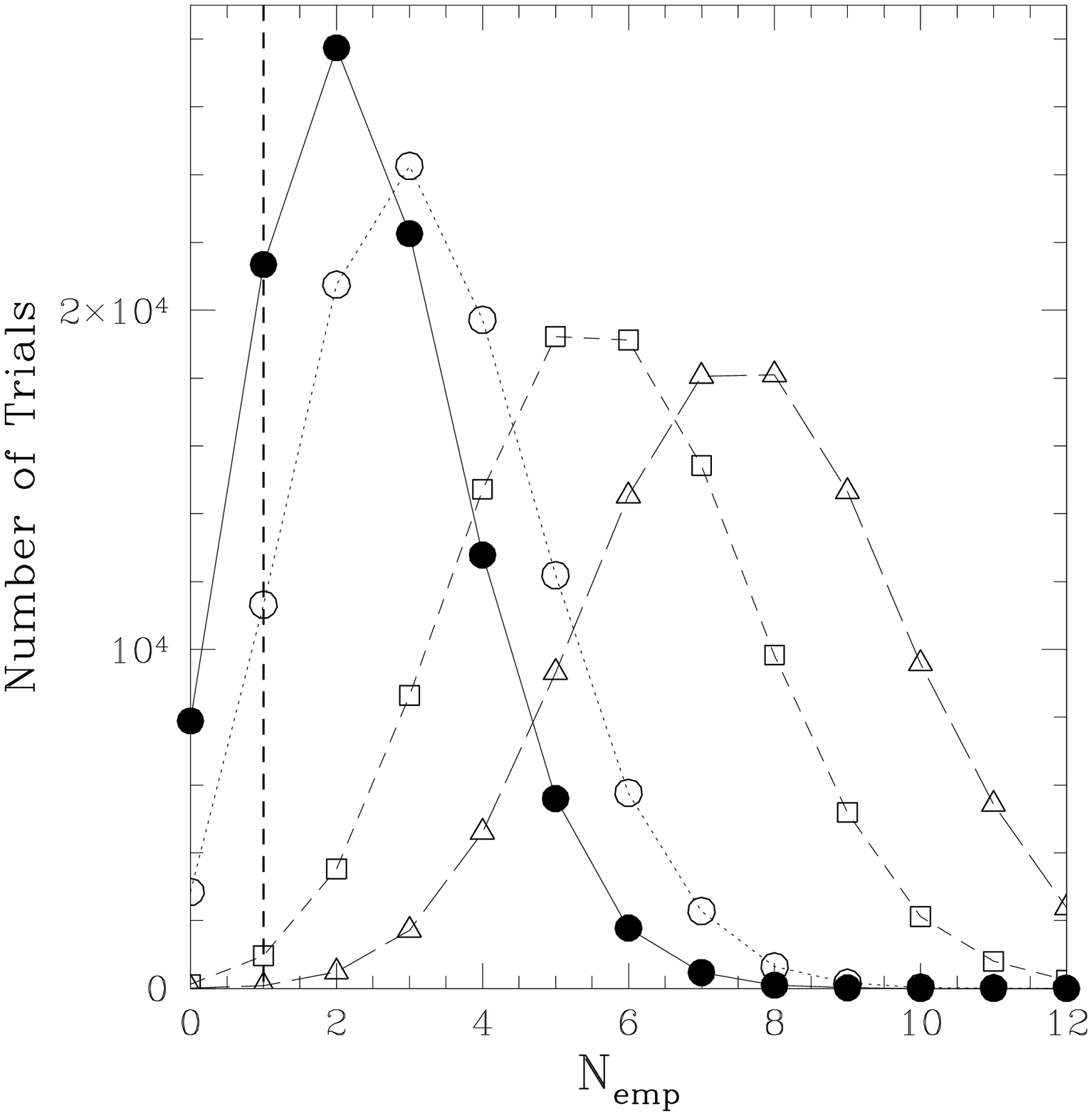}
\caption{{\bf Left} panel: fraction of trials from Montecarlo
simulations that match observations (i.e. 1 cluster in the ``empty''
region) as a function of $f$ (see text), assuming ${\cal N}$= 17 (solid line),
19 (dotted line) and 20 (dashed line) (Table 3).
{\bf Right} panel: distributions of trials from Montecarlo
simulations as a function of the number of clusters in the
empty region, ${N}_{emp}$, assuming $f$=0.125 (solid line
and filled circles), 0.17 (dotted line and empty circles),
0.3 (dashed line and empty squares), 0.4 (long-dashed line and 
empty triangles). $\tau = 3.5 f$ and $=3.5 f/2$ Gyrs if one assumes
that clusters cross the ``empty'' region one or two times, respectively.
The vertical dashed line marks 1 cluster in the ``empty'' region.
}
\end{center}
\end{figure*}

\noindent
On the other hand, clusters with radio upper limits, typically
more relaxed than radio halo clusters (Venturi et al. 2008),
must have experienced the last merger at earlier epochs : after the last
merger they already had sufficient time for suppression of the
synchrotron emission and consequently
they should be the ``oldest'' systems in the GMRT sample.

\noindent
Clusters in the ``empty'' region may be (a) ``intermediate'' systems 
at late merging phases, where synchrotron emission is being suppressed,
or (b) the ``very young'' systems in the very early phases of a merging
activity, where synchrotron emission is increasing.
One cluster in the GMRT sample
is found in the ``empty'' region, this is the merging cluster RXJ1314
that hosts a small-scale radio halo and 2 radio relics (Feretti et al.
2005; Venturi et al. 2007) and consequently
it likely belongs to the latter class (b) of galaxy clusters (although
we cannot exclude that it is an ``intermediate'' system).

\section{Constraining the evolution of radio halos}

In the case that we admit that radio halos are transient phenomena
connected with cluster-merging phases, 
the ``emptiness'' of the region between radio halos and ``radio quiet'' 
clusters in the $P_{1.4}$--$L_X$ diagram can be used to constrain 
the time-scale of the evolution (suppression and amplification) 
of the synchrotron emission in these clusters (Brunetti et al. 2007).
Indeed the significant lack of clusters in this region suggests
that this time-scale is much shorter than both the ``life-time''
of clusters in the sample and the period of time clusters spent in the
radio halo stage.

\noindent
By restricting our analysis to the 19 clusters in the GMRT sample
with $L_X \geq 8.5 \times 10^{44}$erg s$^{-1}$,
in which case the radio power of giant radio halos is 1 order of magnitude
larger than radio upper limits, we find 5 giant radio halos (and 2 
mini-halos in cool-core clusters) on the correlation,
11 clusters in the region of the upper limits and only RXJ1314 in the
``empty'' region.
Thus, the time interval 
that clusters may spend crossing this ``empty'' region is
$\approx f \tau_{gc}$, where $f$ is the
fraction of clusters in this region, $1/19$, and $\tau_{gc}$ is the period
of time 
elapsed since the last merger in the case of the ``oldest'' clusters in
our population.
Since the GMRT sample is constituted by massive,
$M > 10^{15}$M$_{\odot}$, galaxy clusters at z=0.2-0.4,
$\tau_{gc}$ is essentially the time-scale between the epoch of
formation of these clusters (z$\approx$0.6-0.7, e.g. Giocoli
et al.~2007) and the most recent
epoch of observation (z=0.2, $\approx 11.2$ Gyr), $\tau_{gc} \approx 3.5$ Gyr.
Consequently, the ``life-time'' of radio halos is $\tau_{rh}=7/19 \tau_{gc}
\approx 1.3$ Gyr,
the time interval that clusters may spend in the ``empty'' region
is $=1/19 \tau_{gc} \approx 180$ Myr, and the corresponding time-scale 
for suppression 
of the cluster-scale synchrotron emission from the level of radio halos 
to that of ``radio quiet'' clusters, $\tau$,
is roughly half of this period, $\tau \approx$ 90 Myrs, considering
that clusters may cross the ``empty'' region two times
(during the amplification and suppression of synchrotron emission, Sect.~3).
We stress that this conclusion holds even if the
two mini-halos, Abell 2390 and Z7160, are excluded from our 
analysis\footnote{Mini-halos in cooling-flow clusters may have a 
different origin with respect to giant radio halos, possibly connected 
with the presence of the cooling flow (Gitti et al.~2002; 
Mazzotta \& Giacintucci 2008; Cassano et al.~2008b).}, 
in this case $\tau_{rh}\approx 1$ Gyr, $f = 1/17$, 
and $\tau \approx$ 200 Myrs.

Obviously the poor statistics allows fairly large uncertainties on
the above numbers.
Consequently, because in Sect.5 we will discuss the implications of 
these constraints 
for the origin of radio halos, it is important to understand whether, due to 
the poor statistics, a significantly larger value of $\tau$ is 
still consistent with the distribution of galaxy clusters in Figure 1.
We use Montecarlo procedures : we assume that clusters spend a fraction 
of time $f_{low}$, $f_{up}$ and $f$ in the region of the upper-limits, 
on the correlation and in the ``empty'' region, respectively.
Then we perform $10^5$ random extractions of $\cal{N}$ synthetic clusters
with probability $f_{low}$, $f_{up}$ and $f$ 
(with $f_{low} + f_{up} + f =1$), and derive the fraction of trials 
where 1 cluster falls in the ``empty'' region.
This is shown in Figure 2 (Left) as a function of $f$ :
as expected the distribution peaks at $f \simeq 0.06$, corresponding
to $\tau \sim 1/2 f \tau_{gc} \approx$ 100 Myr, in which case about
38\% of trials match observations. 
In the less constrained case, $\cal{N}$=17, where the 2 mini-halos in the
GMRT sample are not considered, the fraction of trials that 
match observations falls to only 1 \% and 0.3 \% for $f$=0.33 
($\tau = 0.57$) and $f$=0.39 ($0.68$ Gyrs), respectively (see Table 3, 
where we also report the results obtained by considering different
sub-samples).

This analysis allows us to conclude that values of $\tau$ significantly 
larger than a few tenths of Gyr are very unlikely.
This is also highlithed by Figure 2 (Right) that shows the 
distribution of trials (in the less constrained case, $\cal{N}$=17)
as a function of the number of galaxy
clusters found in the ``empty'' region for different values
of $\tau$.
Larger values of $\tau$ imply an increasing number of
clusters expected in the ``empty'' region, and 
$\tau \geq 0.6$ Gyr can be excluded at $> 99\%$ confidence level.
We note that our conclusion is inconsistent with much larger 
values of the transition time-scale, $\tau \sim$ few Gyrs, as
recently claimed by Kushnir et al.~(2009) that however estimate
the transition time-scale as $\tau \sim 1/3 \tau_{gc}$, $1/3$ being 
the fraction of clusters with radio halos and $\tau_{gc}$ 
taken = 5 Gyr\footnote{This approach indeed would give the life-time of radio 
halos, not the transition time-scale, and it is indeed consistent with our 
estimate of $\tau_{rh}$.}.
On the other hand, our statistical analysis provides 
more quantitative support to previous conclusions
(Brunetti et al.~2007; Brunetti 2008).

For completeness, we also consider the complementary scenario where
clusters cross the ``empty'' region only one time, due to
the suppression of their synchrotron emission \footnote{This might happen
if the increase of the cluster X-ray
luminosity during mergers takes longer times than that of the
synchrotron luminosity, and consequently merging clusters may
approach the range of X-ray luminosities of the GMRT sample
``along'' the correlation.}. In this case we interpret RXJ1314 as
an ``intermediate'' system in early-post merging phase (Sect.~3) that
provides the most conservative approach to constrain $\tau$
(Table 3).
Still also in this conservative approach we conclude that
present data strongly favour values of $\tau$ substantially smaller 
than 1 Gyr ($\tau \geq$ 1 Gyr is excluded at $>99$\%
in our reference case, ${\cal N}=19$, Table 3).

\begin{table*}
\begin{center}
\begin{tabular}{c|c|c|c|c|c|c|c}
\hline
${\cal N}$ & ${N}_{corr}$ & ${N}_{emp}$ & ${N}_{ul}$ & 
$\tau_{10\%}$ & $\tau_{5\%}$ & $\tau_{1\%}$ & $\tau_{0.3 \%}$ \\
   &   &   &    &  (Gyr) & (Gyr) & (Gyr) & (Gyr) \\
\hline
\hline
17 & 5 & 1 & 11 & 0.35 & 0.42 & 0.57 & 0.68 \\
{\bf 19} & {\bf 7} & {\bf 1} & {\bf 11} & {\bf 0.32} & 
{\bf 0.34} & {\bf 0.51} & {\bf 0.60} \\
20 & 7 & 1 & 12 & 0.30 & 0.36 & 0.49 & 0.59 \\
\hline
\hline
17 & 5 & 1 & 11 & 0.69 & 0.84 & 1.14 & 1.35 \\
{\bf 19} & {\bf 7} & {\bf 1} & {\bf 11} & {\bf 0.64} & 
{\bf 0.68} & {\bf 1.01} & {\bf 1.20} \\
20 & 7 & 1 & 12 & 0.60 & 0.73 & 0.97 & 1.17 \\
\hline
\end{tabular}
\end{center}
\caption{Synchrotron dissipation time (in unit of Gyr) in galaxy clusters
that allows to match observations (1 cluster in the `empty'' region)
in a fraction of Montecarlo trials = 10\% (col.5), 5\% (col.6), 1\% 
(col.7) and 0.3 \% (col.8).
Montecarlo simulations are carried out for three configurations: GMRT
clusters with $L_{0.1-2.4}>8.4 \cdot 10^{44}$erg s$^{-1}$ excluding
the two mini-halos (first line), GMRT clusters with
$L_{0.1-2.4}>8.4 \cdot 10^{44}$erg s$^{-1}$ (second line),
GMRT clusters with $L_{0.1-2.4}>8 \cdot 10^{44}$erg s$^{-1}$ excluding
the two mini-halos (tirdth line); total number of observed clusters
and their distribution are given in col.~1-4.
We perform $10^5$ random extractions of ${\cal N}$ synthetic clusters
assuming a grid of extraction probability with $f$ in the range 0--1 
(see text).
Upper part gives the case where clusters cross the ``empty'' region
two times, lower part gives the case where clusters cross
this region only one time.
}
\end{table*}

\section{Implications for the origin of giant radio halos}

\subsection{Hadronic models}

Theoretically relativistic protons are expected to be the dominant
non-thermal particle component in galaxy clusters since they have very 
long life-times and remain confined within clusters for an Hubble time 
(e.g. Blasi et al.~2007 and ref. therein). 
Proton-proton (p-p) collisions provide a continuous source of secondary 
products in the ICM, and secondary electrons in turns generate diffuse 
synchrotron emission.

\noindent
Radio halos are found in merging clusters and 
the passage of merger shocks through the ICM may increase the energy 
density of protons (e.g. due to acceleration of these 
protons at merger shocks) enhancing the rate of production of secondary 
electrons and the resulting cluster-scale synchrotron emission.
However, since protons have very long life-times the production rate
of secondary electrons in the ICM would remain unchanged with cosmic time
and the mechanism itself does not allow a suppression of the 
synchrotron emission when clusters become more dynamically relaxed.

\noindent
Clusters of equal masses may experience different formation
histories, yet the global budget of gravitational energy dissipated
at merging and accretion shocks is expected to be similar yielding 
to fairly small differences in terms of energy content of relativistic
protons (Jubelgas et al.~2008). Consequently,
assuming that the ICM is magnetised at $\mu$G level, radio halos generated 
by secondary emission are expected {\it long-living} and common;
also, some trend between their radio power and the X-ray luminosity or 
temperature of the hosting clusters is expected
(Dolag \& Ensslin 2000; Miniati et al. 2001; Dolag 2006; Pfrommer et al. 2008).
Due the huge uncertainties in the physics of shock 
acceleration one may easily believe that large variations of the 
content of relativistic protons are possible among clusters with similar mass 
(temperature ..), however there is no reason to expect a {\it bi-modality} 
in the cluster synchrotron emission.

Consequently, to explain the separation between radio-halo 
and ``radio quiet'' clusters and the merger--halo connection
the magnetic field should play a major role and 
it must be admitted that merging clusters, hosting
radio halos, have larger magnetic fields and that this excess
in magnetic field is dissipated when clusters become ``radio quiet'' and 
dynamically relaxed (Brunetti et al.~2007, 2008; Kushnir et al.~2009).

\noindent
Synchrotron emission in hadronic models scales as
(e.g. Dolag \& Ensslin 2000):

\begin{equation}
\epsilon_{syn} \propto {{ B^{1+\alpha} }\over{B^2 + B_{cmb}^2}}
\,\,\, ,
\end{equation}

\noindent
where $B_{cmb} =3.2 (1+z)^2 \, \mu$G is the equivalent field due to inverse 
Compton scattering of
Cosmic Microwave Background photons and $\alpha \sim 1.3$ is the synchrotron 
spectral index of radio halos (e.g. Ferrari et al.~2008). Thus 
to explain a suppression $\geq$10 in terms of synchrotron
emission (Figure 1) the ratio between the magnetic 
fields in radio halos,
$B + \delta B$, and that in ``radio quiet'' clusters, $B$, must
be :

\begin{equation}
( {{ B + \delta B }\over{ B }} )^{\alpha-1}
{{ 1 + ( {{B_{cmb} }\over{B }} )^2 }\over
{1 + ( {{B_{cmb} }\over{B + \delta B}} )^2 }}
\geq 10
\label{eq:gap}
\end{equation}

\noindent
The ratio between the magnetic field energy densities in the two cluster
populations, $\omega_{rh}/\omega_{rq} \propto (1 + \delta B /B)^2$,
from Eq.~3 is shown in Figure 3 for $z\approx 0.25$, typical of
GMRT clusters.
In the case $B + \delta B << B_{cmb}$, hadronic models
must admit that the energy density of the magnetic field in
``radio quiet'' clusters is $\geq$10 times smaller than that
in radio halos, and even larger ratios
must be admitted in the case $B + \delta B >> B_{cmb}$.

\begin{figure}
\begin{center}
\includegraphics[width=7cm]{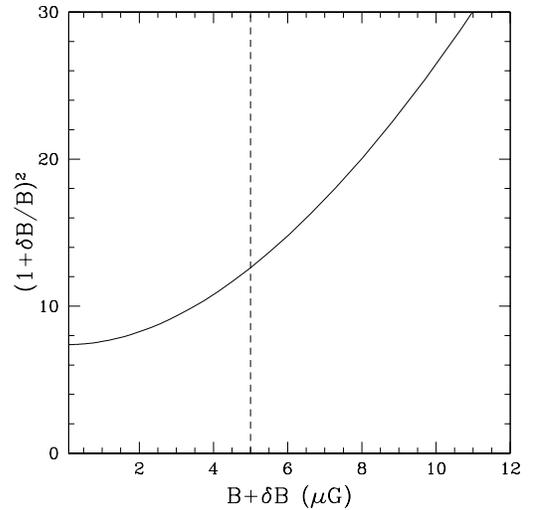}
\caption{Lower limit to the ratio between the energy density of 
the magnetic field
in radio halos and in ``radio quiet'' clusters as a function of
the magnetic field strength in radio halos.
The vertical dashed line marks the value of the equivalent field
of the Cosmic Microwave Background photons assuming z=0.25.}
\end{center}
\end{figure}

\begin{figure}
\begin{center}
\includegraphics[width=7cm]{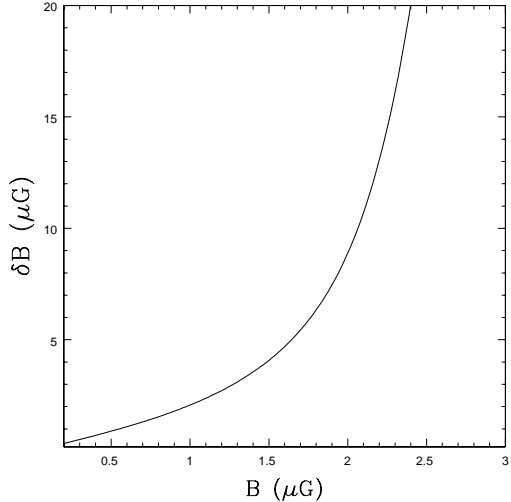}
\caption{The intensity of the small-scale magnetic field in radio
halos that must be dissipated to match
observations is reported as a function of the strength of the
large-scale magnetic field in ``radio quiet'' (and radio halo)
clusters.}
\end{center}
\end{figure}

Theoretically we might admit that 
the magnetic field is amplified in the ICM by turbulence
generated in cluster mergers (Dolag et al.~1999, 2005; Ryu et al.~2008),
and later dissipated since turbulent magnetic fields can decay.

\noindent
On the other hand, to our knowledge, studies of Faraday Rotation in galaxy
clusters do not find any statistical difference, in terms of 
energy density of the large scale (10-100 kpc coherent scales) magnetic 
field, between clusters hosting radio halos and clusters without Mpc-scale 
radio emission (e.g. Carilli \& Taylor 2002 \footnote{We would also point
out that in some cases the magnetic field in ``radio quiet'' clusters,
e.g. A119, is larger than that of radio halo clusters, A2255, with similar 
X-ray luminosity (Murgia et al.~2004, Govoni et al.~2005).}).

Most important, dissipation of this magnetic field is expected to take 
long time.
Even if we simply consider the case where the field is dissipated through 
the decay of cluster-MHD turbulence, the energy density of the rms field 
decreases only (about) linearly with the eddy turnover time-scale. 
This requires several eddy turnover times, $\approx$ a few Gyr, to gradually
dissipate the bulk (i.e. 80-90 \%) of the energy density of the field in
the ICM (Subramanian et al.~2006), that indeed also explains why 
few $\mu$G--fields are common in galaxy clusters. 
Consequently the dissipation time-scale of the magnetic field
is inconsistent with (larger than) that of the suppression of
the cluster-scale synchrotron emission inferred from the statistical 
analysis in previous Section.

\noindent
This conclusion is based on the scenario, following Subramanian et al.
(2006), that the magnetic field in the ICM is amplified by cluster turbulence
generated on large scale, $\sim$150-300 kpc, in which case the thickness
of magnetic filaments is expected to be $\sim$20-40 kpc.
This is supported by Faraday Rotation measurements of extended sources
in clusters that allow to observe ordering scales of the magnetic field
$\sim$10-40 kpc (Clarke et al.~2001; Guidetti et al.~2008) and that
indicate, at least in some cases, that the power spectrum of the 
magnetic field extends to very large scales, 100-500 kpc (e.g. Murgia
et al.~2004).
On the other hand our understanding
of the origin and of the properties of magnetic field in galaxy clusters
is still poor and leaves space to large uncertainties.
Indeed, faster dissipation of the magnetic field in galaxy clusters,
$\tau \approx$ several 100 Myr, may happen in the case that
the magnetic field in excess, $\delta B$, in clusters hosting
radio halos is associated with a field component on smaller scales.
The value of the small scale field, $\delta B$, necessary to account for 
the difference between the synchrotron emission in radio halo 
and ``radio quiet'' clusters can be obtained from Eq.~3 and is reported 
in Figure 4 as a function of the large scale magnetic field, $B$.
Figure 4 clearly highlights the drawbacks of this hypothesis : first of all
the small scale field must be energetically dominant with respect
to that on larger scales (see also Figure 3), in addition if the
large scale field is $\geq 1.5-2 \mu$G level (consistent with present
RM studies) then $\delta B$ would be extremely large, $\geq$ 10 $\mu$G.
Since $\delta B$ must be dissipated in a few tenths of Gyr, we note that
this would imply a magnetic-energy dissipation rate 
in a Mpc$^3$ region $\geq 10^{46} (\tau\, /0.5 {\rm Gyr})^{-1}$ erg/s,
e.g. larger than the bolometric X-ray emission of clusters themselves.

\subsection{Turbulent acceleration of particles}

MHD turbulence generated during cluster mergers may 
accelerate relativistic particles (e.g. Brunetti et al. 2008).
Even without considering the dissipation (or amplification) of
the magnetic field in the ICM, the finite dissipation time-scale
of turbulence implies that giant radio halos should be found in
merging-clusters where turbulence is still generated and
must be extremely rare in more relaxed clusters (e.g., Cassano
\& Brunetti 2005).

\begin{figure}
\begin{center}
\includegraphics[width=7cm]{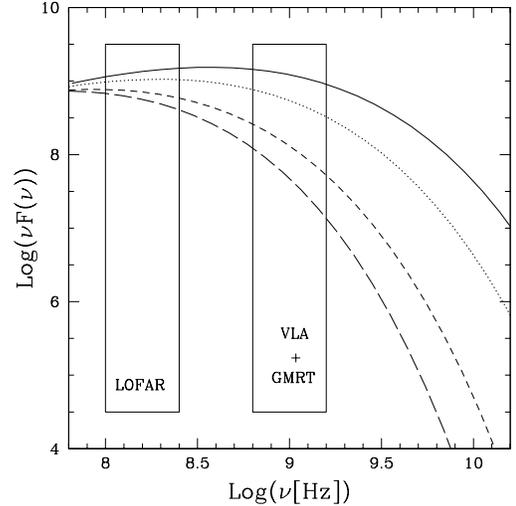}
\caption{Example of the synchrotron emitted power (arbitrary units) as a 
function of frequency for different energy densities of turbulence
(magnetosonic waves) : 30 \% (solid line), 25 \% (dotted line),
20 \% (dashed line), 15 \% (long-dashed line) of the thermal energy density.
In the calculations we adopt a homogeneus model with magnetic 
field strength = 3 $\mu$G, number density of the thermal plasma
= 10$^{-3}$cm$^{-3}$ and temperature T=$7 \cdot 10^7$K.
Frequency ranges of interest for GMRT, VLA and LOFAR are also marked.}
\end{center}
\end{figure}

As soon as large scale turbulence in the ICM reaches smaller, resonant, 
scales (via cascading or induced plasma instabilities, e.g. 
Brunetti et al. 2004, Lazarian \& Beresnyak 2006, Brunetti \& Lazarian 2007),
particles are accelerated and generate synchrotron emission.
In the case of radio halos emitting at GHz frequencies 
the acceleration process should be relatively efficient and 
particles get accelerated to the energies necessary to produce synchrotron
GHz--emission within a time-scale smaller than a couple of cooling times of 
these electrons, that is $\approx 100$ Myrs.
Although the large uncertainties in the way large scale
turbulence is generated in the ICM during cluster mergers, it is likely 
that the process persists for a few crossing times of the cluster-core 
regions, that is fairly consistent with a radio halo life-time $\tau_{rh}
\sim 1$ Gyr as derived in Section 4.

\noindent
Most important, the cooling time of the emitting electrons
is smaller than (or comparable to) the cascading time-scale of 
the large-scale turbulence implying that the evolution of the synchrotron 
power depends very much on the level of MHD turbulence in the ICM (e.g.,
Cassano \& Brunetti 2005; Brunetti \& Lazarian 2007).
Consequently, if we simply assume that the injection of MHD turbulence
is suppressed ``instantaneously'' at a given time (eg. at late
merging-phase), then also the synchrotron emission at higher 
radio frequencies is suppressed, falling below the detection limit of 
radio observations, as soon as the energy density of turbulence starts 
decreasing. 
This is shown in Figure 5 where the synchrotron spectrum
from turbulent accelerated electrons is reported
for different energy density of the MHD turbulence.
A reduction of the turbulent energy density of a factor 2 happens within
about 1 eddy turnover time of the large scale turbulence, that is a few 
times 100 Myrs, and this is sufficient to suppress the synchrotron emission 
at higher, GHz, frequencies by about 1 order of magntitude.

Consequently cluster {\it bi-modality} in this scanario may be expected 
because the transition between radio halos and ``radio quiet'' 
clusters in the $P_{1.4}$--$L_X$ diagram is expected to be fairly fast   
(Brunetti et al. 2007, 08) provided that the 
acceleration process we are looking in these sources is not
very efficient, being just enough to generate radio halos emitting 
at a few GHz frequencies.
On the other way round, we might say that the observed cluster
{\it bi-modality} constrains the efficiency of the particle
acceleration process in radio halos.
Interestingly, a relatively inefficient electron acceleration process 
in radio halos is in line with the steep spectrum observed in these sources
and, most important, with the presence of a spectral steepening at 
higher frequencies discovered in a few halos (e.g., Thierbach et al. 2003,
Brunetti et al. 2008, Dallacasa et al. 2009).

\noindent
Also, Figure 5 suggests that under these conditions
the suppression of synchrotron 
emission that follows the dissipation of MHD turbulence is more 
efficient at higher frequencies and thus cluster {\it bi-modality} 
is expected to be less 
pronounced in considering the synchrotron emission of galaxy clusters
at lower frequencies. This is a clear expectation of the scenario that 
can be tested with future observations of samples of galaxy clusters 
at 100--200 MHz
that may be carried out with LOFAR in a couple of years.

\section{Conclusion}

The ``GMRT Radio Halo Survey'' allows to study the statistics
of radio halos in a complete sample of X-ray luminous
galaxy clusters (Venturi et al.~2008).
The high sensitivity of the radio observations at the GMRT allows to
unveil a cluster radio {\it bi-modality} with ``radio quiet''
clusters well separated from the region of the
$P_{1.4}$--$L_X$ correlation defined by giant radio halos 
(Brunetti et al.~2007 and Figure 1).

In the framework of the hierarchical model galaxy clusters are
expected to evolve in the $P_{1.4}$--$L_X$ plane, in which case 
the distribution of GMRT clusters in Figure 1 results from a statistical
sampling of this evolution.
The connection between radio halos and cluster mergers suggests that 
the Mpc-scale synchrotron emission in galaxy clusters is amplified during 
these mergers and then suppressed when clusters become more 
dynamically relaxed.
The separation between radio halo and ``radio quiet'' clusters in
Figure 1, and the rarity of galaxy clusters with intermediate radio
power suggests that the processes of amplification and suppression
of the synchrotron emission takes place in a relatively short
time-scale.

The time-scale of the 
evolution from radio halos to ``radio quiet''
clusters (and vice versa) provides a novel tool to constrain models 
proposed for the origin of radio halos, namely the re-acceleration 
and hadronic model. 
In the former case the acceleration and cooling of relativistic electrons 
drive the level of the Mpc-scale synchrotron emission from clusters, 
while in the latter case the transition between radio halo and 
``radio quiet'' clusters must be due to the amplification and 
dissipation of the magnetic field in the ICM.

\noindent
We carried out statistical analysis of the cluster radio {\it bi-modality}
in Figure 1 and, although the still poor statistics, 
show that the suppression of the cluster-scale 
synchrotron emission must happen in a fairly short time-scale, a few
100 Myrs, whereas longer time-scales, Gyr, are not
consistent with present data.
This short transition time-scale can be potentially reconciled with the 
hypothesis that the emitting electrons are accelerated by cluster-scale 
turbulence, in which case the synchrotron radiation emitted at GHz 
frequencies may rapidly decrease as a consequence of the dissipation of 
a sizeable fraction of that turbulence.
In this case, however, we also claim that a less pronounced {\it bi-modality} 
is expected in the case of cluster samples observed at lower
radio frequencies, that may be tested by future LOFAR and LWA observations.

\noindent
On the other hand, it is more difficult to reconcile a short transition 
time-scale in the case that the unique source of emitting electrons
is provided by p-p collisions (hadronic models).
In this case the dissipation of the cluster
magnetic field that suppresses the synchrotron emission would take
longer periods of time.
In principle this difficulty could be considerably alleviated in the 
case that the energy density of the magnetic field in radio halos
is dominated by that of small scale field. However, we would
come into the untenable scenario in which a very strong, transient 
magnetic field (small scale) component, is present in the ICM. 
Future studies of source-depolarization in cluster radio sources will 
also help in constraining the level of the small-scale field component 
in the ICM.

\noindent
We stress that our constraints come from the conservative (and
simplified) assumption that the injection of turbulence (as well as the 
amplification of the magnetic field) switches off at the same time 
across the radio halo, Mpc$^3$, region.
In reality, depending on the way turbulence and large scale magnetic fields
are generated in the ICM, the suppression and amplification of the 
synchrotron emission could start at different times in different 
parts of this volume.
The most important consequence of that is an expected scatter in
the correlation rather than in the way clusters become ``radio quiet'',
since clusters are expected to start moving 
across the transition region only when synchrotron
is suppressed across a substantial fraction of the radio halo's volume.
Yet, overall this goes into the direction to strengthen our conclusion 
that an efficient process 
to suppress the cluster-scale synchrotron emission in galaxy clusters is
necessary to explain observations.

\noindent
Giant radio halos prove complex physical processes where a fraction of
the gravitational energy dissipated during cluster-mergers is channelled 
into the acceleration of relativistic particles.
The correlation traced by halos in Figure 1 and its intrinsic scatter,
together with the distribution of clusters in the $P_{1.4}$--$L_X$ plane,    
provide novel tools to hopefully constrain the complex physics of turbulence 
and magnetic fields in the ICM and their interplay with the process of 
cluster formation.
The deep surveys at low frequencies with LOFAR and LWA will be crucial 
to overcome present uncertainties due to the still poor statistics
allowing a major step forward in understanding the origin and
evolution of the cluster-scale synchrotron emission. 
Remarkably, as discussed in this paper, the scatter of the correlation 
and the distribution of clusters in the $P_{1.4}$--$L_X$ plane are expected
to depend on the observing radio frequency (Sect.~5.2, Figure 5), and 
consequently deep complementary
follow ups at intermediate and higher frequencies (GMRT, 
eVLA, SKA) will also be crucial.

\begin{acknowledgements} 
This research is partially funded by INAF and ASI through
grants PRIN-INAF2007 and ASI-INAF I/088/06/0.
GB would like to acknowledge P.Blasi, T.Ensslin and D.Kushnir for
stimulating discussions.
\end{acknowledgements}

\end{document}